%% file: recsys2021goal.tex
\documentclass[sigconf]{acmart}
\usepackage{booktabs} 
\include{mypackages}

\usepackage{bm}

\setcopyright{acmcopyright}

\newcommand{\cj}[1]{\textcolor{blue}{\bf\small [#1 --cj]}}

\copyrightyear{2021}
\acmYear{2021}
\setcopyright{acmcopyright}\acmConference[RecSys '21]{Fifteenth ACM Conference on Recommender Systems}{September 27-October 1, 2021}{Amsterdam, Netherlands}
\acmBooktitle{Fifteenth ACM Conference on Recommender Systems (RecSys '21), September 27-October 1, 2021, Amsterdam, Netherlands}
\acmPrice{15.00}
\acmDOI{10.1145/3460231.3474260}
\acmISBN{978-1-4503-8458-2/21/09}

\begin{document}

\title[Learning to Represent Human Motives for Goal-directed Web Browsing]{Learning to Represent Human Motives for Goal-directed Web Browsing}




\author{Jyun-Yu Jiang}
\authornote{Work done while at Microsoft.}
\email{jyunyu@cs.ucla.edu}
\affiliation{%
  \institution{University of California, Los Angeles}
  \country{USA}
}

\author{Chia-Jung Lee}
\authornotemark[1]
\email{cjlee@amazon.com}
\affiliation{%
  \institution{Amazon Inc.}
  \country{USA}
}

\author{Longqi Yang}
\email{loy@microsoft.com}
\affiliation{%
  \institution{Microsoft}
  \country{USA}
}

\author{Bahareh Sarrafzadeh}
\email{bahar.sarrafzadeh@microsoft.com}
\affiliation{%
  \institution{Microsoft}
  \country{USA}
}

\author{Brent Hecht}
\email{brhecht@microsoft.com}
\affiliation{%
  \institution{Microsoft}
  \country{USA}
}

\author{Jaime Teevan}
\email{teevan@microsoft.com}
\affiliation{%
  \institution{Microsoft}
  \country{USA}
}


\renewcommand{\shortauthors}{J.-Y. Jiang et al.}
\begin{abstract}
\input{sections/s0_abstract}

\end{abstract}

%
%


\maketitle

\input{sections/Introduction}

\input{sections/s2_relatedwork}

\input{sections/s3_goalspace}

\input{sections/s4_model}
\input{sections/s5_experiments}

\input{sections/s6_analysis}
\input{sections/s6_conclusion}


{
\bibliographystyle{ACM-Reference-Format}
\bibliography{recsys2021goal} 
}
\end{document}

%% file: mypackages.tex
\usepackage[ruled,vlined,linesnumbered]{algorithm2e}
\usepackage{subcaption}
\usepackage{bbm}
\usepackage{bm}
\usepackage{enumitem}
\usepackage{multirow}
\usepackage{amsmath}
\usepackage{amsfonts} 
\usepackage{graphicx}
\usepackage{wrapfig}
\usepackage{titlesec}
\titlespacing*{\section}{0pt}{*0.6}{*0.3}
\titlespacing*{\subsection}{0pt}{*0.6}{*0.3}
\titlespacing*{\subsubsection}{0pt}{*0.6}{*0.3}
\usepackage{mathtools}
\usepackage{flushend}
\usepackage{dsfont}

\usepackage{cleveref}
\usepackage{xspace}

\crefformat{footnote}{#2\footnotemark[#1]#3}

\newcommand{\mysection}[1]{\vspace{3pt}\noindent \textbf{#1.}\xspace}

\usepackage{array}
\newcommand{\PreserveBackslash}[1]{\let\temp=\\#1\let\\=\temp}
\newcolumntype{C}[1]{>{\PreserveBackslash\centering}p{#1}}
\newcolumntype{R}[1]{>{\PreserveBackslash\raggedleft}p{#1}}
\newcolumntype{L}[1]{>{\PreserveBackslash\raggedright}p{#1}}

\newcommand{\modelfullname}{\underline{Go}al-directed \underline{We}b \underline{B}rowsing\xspace}
\newcommand{\modelname}{GoWeB\xspace}

\newcommand{\argmax}[1]{\underset{#1}{\operatorname{argmax}}~}

\newcommand{\new}[1]{\xspace {\color{black}#1}}

%% file: sections/s0_abstract.tex
Motives or goals are recognized in psychology literature as the most fundamental drive that explains and predicts why people do what they do, including when they browse the web. 
Although providing enormous value, these higher-ordered goals are often unobserved, and little is known about how to leverage such goals to assist people's browsing activities.  
This paper proposes to take a new approach to address this problem, which is fulfilled through a novel neural framework, \modelfullname~(\modelname). 
We adopt a psychologically-sound taxonomy of higher-ordered goals and learn to build their representations in a structure-preserving manner.
Then we incorporate the resulting representations for enhancing the experiences of common activities people perform on the web.
Experiments on large-scale data from Microsoft Edge web browser show that \modelname significantly outperforms competitive baselines for in-session web page recommendation, re-visitation classification, and goal-based web page grouping.
A follow-up analysis further characterizes how the variety of human motives can affect the difference observed in human behavioral patterns.

%% file: sections/Introduction.tex
\section{Introduction}

Constructs such as motives and goals are recognized in the psychology literature as the fundamental forces that guide human behavior.
While a variety of interpretations of goals exist, there is consensus that these constructs regulate controlled cognitive processes such as planning and resource allocation, and direct behavioral sequences intended to enact specific performances~\cite{Chulef1993}.
When people browse the web, which is one important behavior common in modern lives, goals also underlie why they do what they do. 
For example, they may buy products to work towards their fitness goals, plan trips to stay connected with friends and family, conduct business to pursue career success, or do research or seek health advice in support of their well-being.
Understanding peoples’ goals not only helps answer the ``why'' questions about their activities \cite{Faaborg2006AGW, smith2010ui}, but also signifies where the opportunities sit for potential improvements.

Although goals and motives are fundamental to browsing behaviors, they stay unobserved and cannot be easily inferred due to the variety and complexity of activities prevalent on modern web browsers.
Moreover, the question of how goals and motives can be utilized to assist browsing remains unanswered.
To bridge the gap, we propose to adopt the well-established bodies of psychological theories \cite{Moskowitz2009ThePO, Barua2014longtermgoals, Chulef2001AHT} regarding human motives to the need for browser-centric activities. 
For instance, the motive of \textit{staying physically healthy} may energize a person to consider \textit{purchasing a rowing machine} or \textit{to learn nutritious ingredients} via a number of web interactions. 
As a result, a better understanding of human motives can guide how we might enhance browsing experiences pertaining to people's needs.

Previous research on supporting user goals on the web has primarily sought to understand and categorize a person's intent manifested through search queries.
In their seminal studies, \citet{broder2002searchtaxonomy} and \citet{rose2004usergoal} classified the goal of a user query into one of three categories: navigational, informational, and transactional.
Successive work refined intent categorization by introducing purchase, sell or job search intents~\cite{dai2006detecting, li2008learning}, as well as examined the variability of intents across different users~\cite{teevan2008personalize}. 
The main objective of research in this field is often to inform and improve the search process~\cite{broder2002searchtaxonomy,CARUCCIO2015userintent,rose2004usergoal,Uichin2005}, and the characteristics of user goals tend to be task-oriented or situation-specific.
In contrast to that, we emphasize incorporating the fundamental human goals into the context of web browsing, since they are recognized to be key to the core values in people's lives \cite{Barua2014longtermgoals} (e.g. health, well-being, sustainability, and learning). 

In this paper, we focus on tackling two research agendas toward our goal-directed vision, namely, how to represent influential human goals\footnote{Unless otherwise mentioned, we use human motives and goals interchangeably in this paper.} and how to leverage the learned representations to provide assistance in browsing sessions.
We propose a unified neural framework, \modelfullname (\modelname), for both objectives laid out in two phases.
The first phase concerns human motives.  
Psychology research has developed decades of knowledge informing which human goals are considered influential through theoretical and empirical evidence. Inspired by the findings, \modelname uses these goals as the backbone and builds on the top to learn intrinsic goal representations following a distance-based reconstruction objective to retain inherent hierarchical structures\footnote{For example, in \citet{Chulef2001AHT}'s human goal taxonomy, the goals \textit{be affectionate} and \textit{share feelings} are both decedents of a parent goal \textit{friendship}.}.  
\modelname further devises a neural goal estimator, which learns to transfer the knowledge entailed in these goal representations to estimating people's motives for visiting web pages.
The second phase focuses on enhancing web experiences with what we have learned. Specifically, \modelname integrates the goal estimator with modern language models~\cite{devlin2019bert} to assist three common interactions or needs people have on the web.
We consider    
an in-session web page recommender that recommends web pages to help people advance their goals, a web page re-visitation classifier that predicts if a person will revisit a web page in the future due to recurring or unfinished goals, and a web page grouping method that clusters in-session web pages according to underlying goals. The overall framework is illustrated in Figure~\ref{fig:overall}.

We evaluate \modelname using anonymized browsing logs obtained from the Microsoft Edge browser.   
The experimental results show that \modelname consistently outperforms competitive baselines in all three browser-centric applications, suggesting that capturing fundamental human motives can empirically improve the intelligence of web browsers.
To understand the effectiveness, we find that the learned intrinsic goal representations largely preserve the human-curated hierarchical relations. Meanwhile, the goal estimator can effectively predict the goals behind web page visits evaluated based on quantitative and qualitative exercises. 
Our follow-up analysis further characterizes the browsing patterns when people engage in pursuing different goals. 
The results suggest that people may explore multiple directions simultaneously when the categories of session's focal goals are broad (e.g., \textit{Ethics \& Idealism}), while they may revisit web resources in shorter intervals for goals concerning social interactions (e.g., \textit{Friendship}). 

%% file: sections/s2_relatedwork.tex
\section{Related Work}
\label{section:relatedwork}

Studying goals is an important, long-lasting topic across multiple research disciplines. We examine the different notions of goals that have been investigated in the literature. 

\mysection{Task-oriented Goals in Search and Web Browsing}
Prior work has a strong focus on studying search goals as they contribute to a critical segment of web browser usage. 
The mainstream work in this space follows \citet{broder2002searchtaxonomy}, and \citet{rose2004usergoal} who classified the goal of user queries into navigational, informational, and/or transactional. \citet{Uichin2005} identified that 60\% of search queries can be associated with informational or navigational goals, and proposed using clicks and anchors to predict those.
Caruccio et al.~\cite{CARUCCIO2015userintent} introduced a lightweight taxonomy that added a handful of sub-classes to informational and transactional queries. Casting as a classification problem, they proposed content and behavioral features to predict which class in the taxonomy a query should belong to using data collected from a custom browser extension. 
For more complex search needs, Jones and Klinkner~\cite{jones2008searchtopics} leveraged human annotations and built predictive models that were trained to segment sequences of user queries into same or different search goals. Among a set of hand-crafted features, they showed that lexical features such as words and characters were identified as the most useful.
Similarly, Law and Zhang \cite{Law2011complexsearch} showed that the ability to decompose a complex search goal into sub-problems (i.e., a set of queries and corresponding search results) can better support a user's information need.

General web browsing behavioral patterns tend to involve higher flexibility and complexity than search. 
\citet{kumar2010browsing} analyzed large-scale commercial logs and showed that page visits related to search comprise only one-sixth of the log sample.
In addition to search, they identified that content consumption (e.g., news, portals, games, verticals, multimedia) comprise about half of online page views, while communication (e.g., email, social networking, forums, blogs, chat) represent about one-third of those. As a result, general web browsing patterns require different types of support. 
For example, past studies have shown that people are often unable to focus on or return to their goals when using a web browser due to the prevalence of distractions on the web ~\cite{aagaard2015drawn, tseng2019overcoming}. Research on Cyberloafing \cite{glassman2015monitor, lim2002way} and task resumption on the web  \cite{gillie1989makes, mcfarlane2002scope, o1995timespace} have shown that individuals experience difficulty when resuming their main task after  external- or self-interruptions \cite{czerwinski2004diary, adler2013self}.

\citet{Faaborg2006AGW} introduced a program-by-example goal-oriented web browser that allows people to customize how they organize and interact with the web. A person who visits a recipe website, for example, may want to know the nutritional information. By explicitly demonstrating how to complete nutrition extraction once, the system could potentially evoke the same human-specified macros on similar websites.
\citet{dix2010web} attempted to understand the goals behind user behavior by connecting actions and user-created personal ontology structures.

\mysection{Higher-ordered Human Goals}
Understanding human goals and motives has long been a central area of research in the psychology literature, as the goals of individuals largely direct the behavior in which they engage~\cite{Klinger1987}.    
Earlier attempts in creating goal taxonomies relied on theoretical viewpoints. McDougall~\cite{McDougall1933} presented a list of 13 instincts while Murray~\cite{Murray1938ExplorationsIP} articulated 44 variables of personality as forces determining behavior.
Chulef et al.~\cite{Chulef2001AHT} took an empirical approach and recruited participants of diverse backgrounds to delve into developing a hierarchical human goal taxonomy based on similarity between goals, providing a concrete and comprehensive structure. 
More recently, Talevich et al.~\cite{Talevich2017TowardAC} iterated on the taxonomy derived in \cite{Chulef2001AHT} with several added classes of human motives.

As our work incorporates long-term goals in browsing sessions to provide goal-directed assistance, we review the classic and state-of-the-art technical methods applicable in the three browser-centric tasks, and conclude with a brief summary comparing previous approaches and ours.

\mysection{Recommendation} The task of web page recommendation \cite{Gksedef2010CombinationOW} aims to predict the information needs of users and provide them with recommendations to facilitate their navigation. 
Classic approaches to recommendation relies on mining user-item relationships through collaborative filtering \cite{reommendcf2009} and matrix factorization \cite{He2017NeuralCF,Rendle2009BPR}. 
Session-based approaches \cite{Tang2018PersonalizedTS} take into account session-level user actions as a sequential stream of events, based on which recommendations are made to users. 
In recent years, large-scale deep neural networks \cite{recommenddl2019} have shown top performing effectiveness in recommender systems, including leveraging GRU \cite{Hidasi2016SessionbasedRW}, self-attention blocks \cite{Kang2018SelfAttentiveSR} and BERT-based architectures \cite{Sun2019BERT4Rec} for session-based recommendation.

\mysection{Revisitation} Web page revisitation behavior is prevalent on the web. \citet{Obendorf2007WebPR} suggested that people may revisit web pages to access the same resource again (e.g., for unfinished goals) or may re-access a resource as they expect changed
content (e.g., for recurring goals, such as new headlines on a news site).
\citet{Adar2008revisit} combined quantitative and qualitative approaches and showed that people revisit web pages with varying speed. 
To help resume previously encountered information, \citet{Dontcheva2006Summarizing} presented a web extension that allows people to extract entities of interest and creates an interactive summary of extracted entities (e.g., hotels that a person considers for a trip).

\mysection{Grouping}
Automatic managing and grouping the vast amount of unorganized web pages can help reduce the cognitive load required when navigating the web. 
A common strategy is to classify web pages \cite{Baykan2011ACS, Qi2009WebPC} by the encapsulated topics; popular class choices include the Open Directory Project, the Yahoo! Directory, categorized URLs from the Wikipedia, etc.
\new{To relax the requirement of pre-defined classes}, clustering techniques have been studied to mine related web pages based on semantics or graph partitioning \cite{Patel2011ARO}. Different from prior studies, we adopt the notion of goals as \new{the basis} to associate web pages \new{for web page grouping}.

In this work, we incorporate long-term goals to facilitate effective web browsing experiences, including recommendation, revisitation and goal-based grouping, which differ significantly from previous perspectives and approaches. We leverage Poincar\'e embeddings~\cite{Nickel2017poincare} to derive the goal representations, which are combined with state-of-the-art transformer-based architectures~\cite{devlin2019bert} to model users' web session behavior.

%% file: sections/s3_goalspace.tex
\section{Representation Learning for Human Motives}
\label{section:goalspace}


\begin{figure}[!t]
    \centering
    \includegraphics[width=\linewidth]{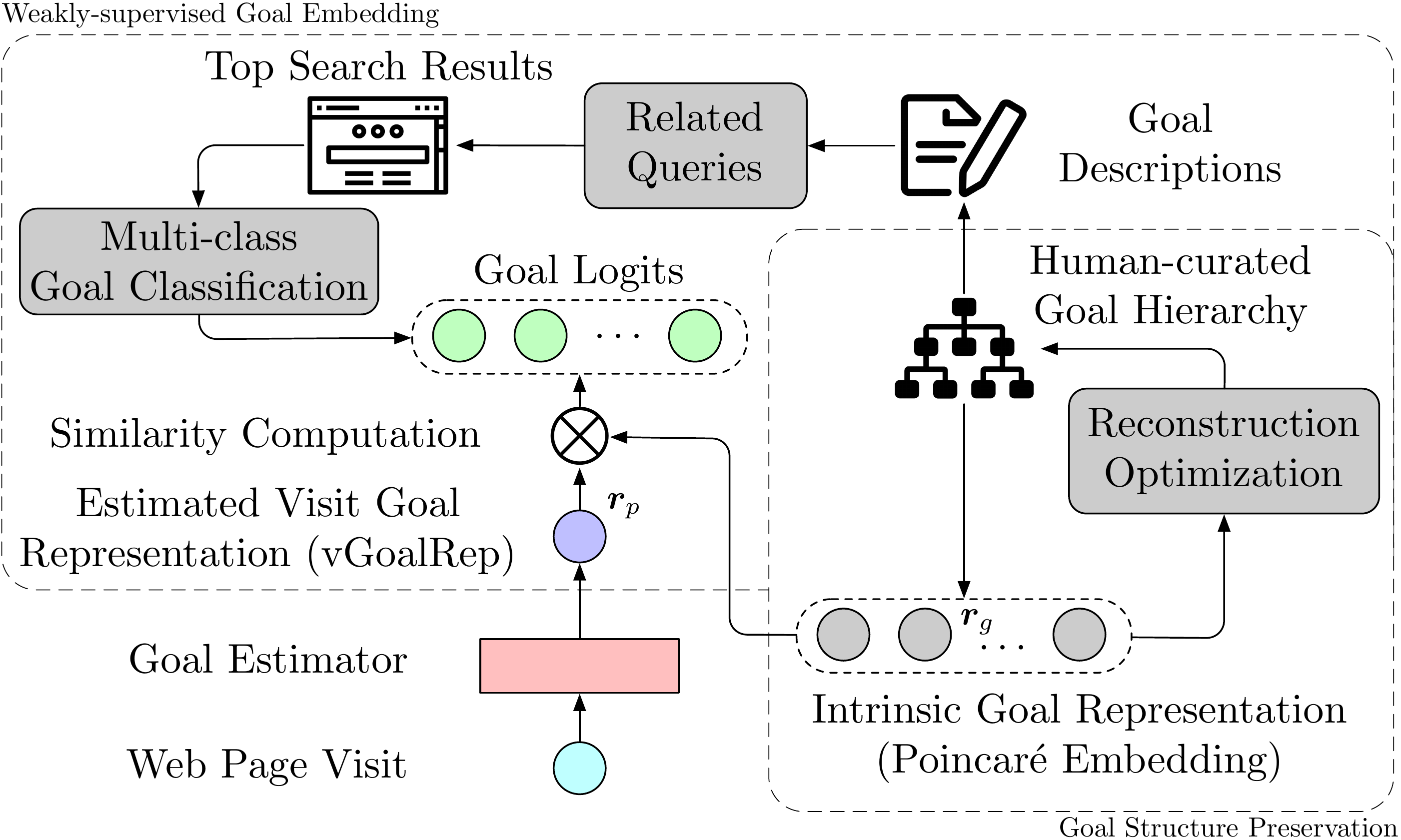}
    \caption{Framework for learning intrinsic goal representations (iGoalRep) and training the goal estimator to predict visit goal representations (vGoalRep).}
    \label{fig:goal_estimator}
\end{figure}

This section presents our approach to learning distributed intrinsic goal representations (iGoalRep $\bm{r}_g$) for human motives, which are grounded by an expert-curated goal taxonomy in the psychology literature.
Moreover, we propose a goal estimator to embed any web page visit $p$ in browsing sessions into the same goal embedding space as $\bm{r}_g$, and we denote them as visit goal representation (vGoalRep) $\bm{r}_{p}$.
To learn the goal estimator, we collect and rely on a weak supervision dataset through the associations between related queries and web pages determined by the Bing search engine.
Figure~\ref{fig:goal_estimator} summarizes the flow of the steps, and we describe the details of each step next.

\subsection{Learning Intrinsic Goal Representations}
\label{section:goalrep}

\begin{figure}[!t]
    \centering
    \includegraphics[width=\linewidth]{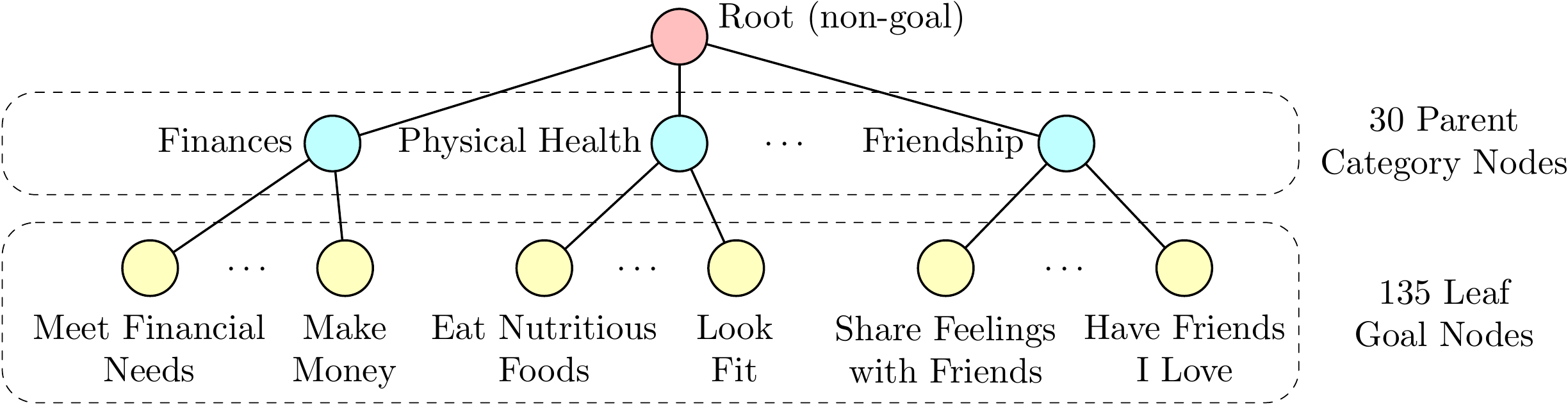}
    \caption{Illustration of the constructed 3-layer goal taxonomy with 166 nodes.}
    \label{fig:goaltax}
\end{figure}

\mysection{Taxonomy of Human Goals}
We leverage the hierarchical human goal taxonomy curated by \citet{Chulef2001AHT} to form the basis of the goal space. 
Specifically, the taxonomy leaf nodes were curated based on an aggregation of psychology literature~\cite{Maslow1970,Murray1938ExplorationsIP,Rokeach1973,Wicker1984} and the responses from participants of diverse background, followed by a similarity-based sorting process to create the hierarchy relationship.
The resulting set consists of 135 leaf goal nodes descended from 30 parent categorical nodes. 
We further introduce a ``non-goal'' node as the root that covers all the categorical nodes. 
Conceptually, the root represents prosaic behavior that might not be tied with specific goals; examples could include random distractions on the web or one-time matters.
Figure~\ref{fig:goaltax} illustrates the final adopted human goal taxonomy consisting of 166 nodes in three layers. 
Formally, $G$ denotes the set of goals in the goal taxonomy and $H$ denotes the hierarchical structure.

\mysection{Representation Learning in Hyperbolic Space}
To induce the structural bias $H$, we propose to learn intrinsic goal representations (iGoalRep) in a hyperbolic space. 
Hyperbolic geometry brings the advantage of learning compact representations that capture both hierarchy and similarity, which, for our case, is desirable as we aim to preserve the hierarchical properties of the goal taxonomy.

Concretely, we adopt Poincar\'e embeddings~\cite{Nickel2017poincare} to derive $d_G$-dimensional iGoalRep $\bm{r}_g$ for each $g \in G$. 
Let $\mathcal{B}^{d_G} = \lbrace \bm{x}\in \mathbb{R}^{d_G} \mid ||\bm{x}|| < 1 \rbrace$ be an open Poincar\'e ball, where $||\cdot||$ indicates the Euclidean norm.
In contrast to Euclidean distance, the hyperbolic distance between any two points $\bm{u}, \bm{v} \in \mathcal{B}^{d_G}$ is given as: 
$$d(\bm{u}, \bm{v}) = \text{arcosh}\left( 1 + 2 \frac{||\bm{u} - \bm{v}||^2}{(1 - ||\bm{u}||^2)(1 - ||\bm{v}||^2)} \right).$$
Subsequent representation learning is then guided by this distance measure where a similar pair of goals should be closer to each other than a remote pair. 


\mysection{Reconstruction Optimization} 
To start learning, we treat the edges in the goal taxonomy as transitive closure and reconstruct the relationship by minimizing the Poincar\'e distances between the embeddings of their endpoints.
Following prior work~\cite{Nickel2017poincare}, we define the loss function $\mathcal{L}_r$ as:
$$\mathcal{L}_r = \sum_{\left(g_u, g_v\right)\in H} \log \frac{e^{-d\left(\bm{r}_{g_u}, \bm{r}_{g_v}\right)}}{\sum_{g_{v^\prime} \in \mathcal{N}(g_u)} e^{-d\left(\bm{r}_{g_u}, \bm{r}_{g_{v^\prime}}\right)}},$$ where $\mathcal{N}(g_u) = \lbrace g_{v^\prime} \mid (g_u, g_{v^\prime}) \notin H  \rbrace \cup \lbrace g_u\rbrace$ denotes a set of  negative examples\footnote{In practice, we randomly sample 50 negative examples for each positive example.} for $g_u$ and $H = \lbrace (g_u, g_v)\rbrace \cup \lbrace(g_v, g_u) \mid g_u, g_v\in G \rbrace$ contains symmetric relationship.
We adopt RSGD~\cite{bonnabel2013stochastic}, a stochastic Riemannian optimization method, to learn goal embeddings within the Poincar\'e ball as in \cite{Nickel2017poincare}.
To do so, we need to rely on the Riemannian manifold structure of the Poincar\'e Ball when minimizing  $\mathcal{L}_r$.
Since Euclidean gradient is not directly applicable in hyperbolic space, rescaling gradients depends on the Riemannian metric tensor $t_x$   equipped in the Poincar\'e ball: $t_x = \left( \frac{2}{1 - ||\bm{x}||} \right)^2 t^E,$ where $\bm{x} \in \mathcal{B}^{d_G}$; $t^E$ denotes the Euclidean metric tensor.

\begin{figure}[!t]
    \centering
    \includegraphics[width=\linewidth]{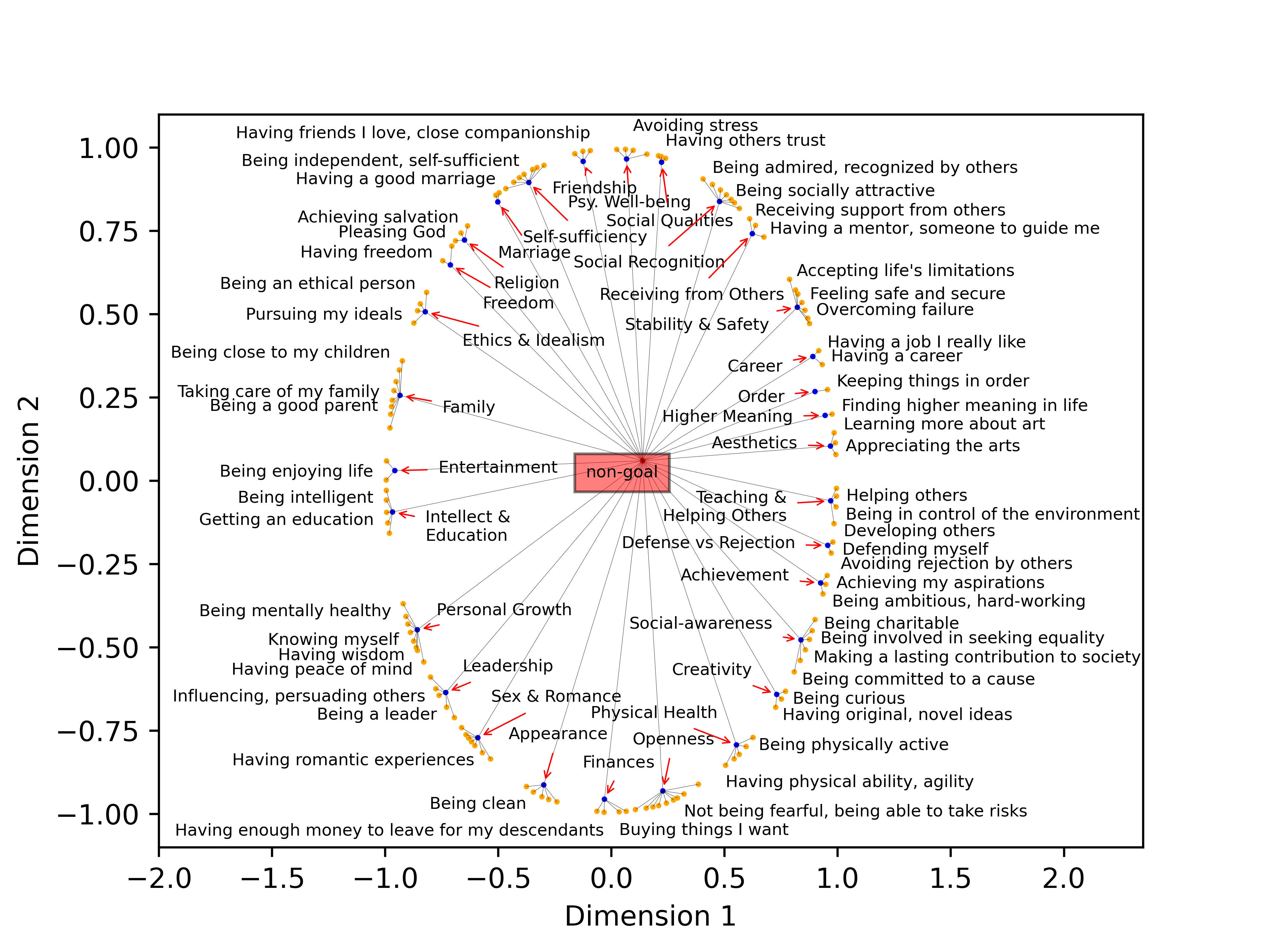}
    \caption{Illustration of learned 2-dimensional Poincar\'e goal representations with the 3-layer goal taxonomy from \cite{Chulef2001AHT}. The center red node indicates the concept of ``non-goal,'' while the blue nodes represent goal categories that cover more specific goals presented as orange nodes.}
    \label{fig:poincareplot}
\end{figure}

Figure~\ref{fig:poincareplot} visualizes the resulting iGoalRep $r_g$ with two dimension\footnote{We select dim=2 for visualization convenience here; in the actual experiments, dim=64 is used as described in Section~\ref{section:exp}.} learned from the goal taxonomy. The results demonstrate that the learned embeddings preserve the desired hierarchical properties and align with the original goal taxonomy curated by human.

\subsection{Estimating Visit Goal Representations (vGoalRep)}
\label{section:goalestimator}

When a person visits a web page, we aim to predict a visit goal representation (vGoalRep) $\bm{r}_p$ such that $\bm{r}_p$ can best reflect which goals may have driven the visit.   
To do so, we first build a weak supervision data collection that provide the association information of what the most probable goals are when web pages are being visited. Then, a parameterized goal estimator is introduced to predict $\bm{r}_p$ based on the weak labels according to a multi-class classification objective.

\mysection{Weak Supervision Data Collection}
Inspired by past work \cite{STROHMAIER201263} that demonstrated search engine queries
can contain explicit user objectives (e.g., get rid of belly fat), 
we first reify every goal in the goal taxonomy with a set of seed queries. 
For example, we manually generate queries such as ``how to be charismatic'' and ``how to meet new friends'' to elicit the high-ordered goal of \textit{being likeable, making friends, drawing others near}.
Then we expand the seed query set with related queries by requesting publicly available Microsoft Bing search API. 
In total, 300 seed queries and 1,932 related queries represent 165 defined human goals. 
The last step involves querying the same API using the enlarged query set together with the original goals; we then keep the top 5 returned web pages as the weakly-positive instances for the corresponding goal. 
For the concept of ``non-goal,'' i.e., the root of the goal taxonomy, we randomly select 1,000 web page visits from general web browsing logs as ``negative examples''. 
Here we denote this data collection as $\bm{D}_{weak}$.

\mysection{Goal Estimator Construction}
Next, we aim to devise a goal estimator that can predict vGoalRep $\bm{r}_p$ by projecting each web page $p$ to the same goal embedding space in which $r_g$ exists.
A web page visit $p$ is commonly characterized by the semantics conveyed via its textual content~$c_p$, as well as the website~$h_p$ where it is hosted (e.g., pages one the same site often demonstrate higher associations).
To build an effective goal estimator, we take as input the two sources of information, $p=$~($h_p, c_p$), and estimate vGoalRep $\bm{r}_p \in \mathbb{R}^{d_G}$ by:
$$ \bm{r}_p = \mathcal{F}_G([\text{emb}^G_{\text{host}}(h_p);\text{BERT}^G(c_p)]),$$
where $\text{emb}^G_{\text{host}}(h_p)$ projects the host to a $d_h$-dimensional embedding space; $\text{BERT}^G(\cdot)$ is a contextualized content encoder, such as BERT~\cite{devlin2019bert} and RoBERTa~\cite{liu2019roberta}; $\mathcal{F}_G(\cdot)$ as a fully-connected hidden layer derives ultimate $d$-dimensional estimated goal embeddings.
The output of the goal estimator, $\bm{r}_p$, can be regarded as a continuous representation of certain underlying motives for visiting a particular web page~$p$.

\mysection{Multi-class Goal Classification}
While the architecture of goal estimator is the same as web page encoder, we learn the parameters using a multi-class goal classification objective guided by $\bm{D}_{weak}$. 
The multi-class classification loss function is defined by categorical cross-entropy~\cite{goodfellow2016deep} as ~$\mathcal{L}_c$:
$\mathcal{L}_c(p) = -\sum_{g_u\in G} \mathds{1}(\hat{g}_p=g_u) \log\left(\sigma_{g_u}\left(\bm{r}_p, G\right) \right)$
where $\mathds{1}(\cdot)$ is an indicator function; $\hat{g}_p$ is the labeled goal of the web page $p$; $\sigma_{g_u}\left(\bm{r}_p, G\right)$ indicates the softmax function calculating the probabilistic distribution over goals in the goal taxonomy as:
$\sigma_{g_u}\left(\bm{r}_p, G\right) = \frac{\bm{r}_p \otimes \bm{r}_{g_u}}{\sum_{g\in G} \bm{r}_p \otimes \bm{r}_g}$
Here $\bm{r}_p \otimes \bm{r}_g$ denotes the similarity score between the web page $p$ and the goal $g$ in the Poincar\'e embedding space. In other words, by treating the Poincar\'e similarity scores as classification logits, $r_p$ is trained to reflect its ground-truth goal embedding $r_g$. Formally: $\bm{r}_p \otimes \bm{r}_g = ||\bm{r}_p||\cdot||\bm{r}_g||\cdot \cos(\theta_{\bm{r}_p}, \theta_{\bm{r}_g})$
where $\theta_{\bm{r}_p}$ and $\theta_{\bm{r}_g}$ indicate the angles of embeddings in the hyperbolic coordinate system.

A final remark to note is that we could derive the most probable goal $g_p$ associated with a page $p$ based on the learned $\bm{r}_p$, by simply calculating the similarity between $\bm{r}_p$ and $\bm{r}_g$: 
$g_p = \argmax{g\in G} \bm{r}_p \otimes \bm{r}_g,$
When designing \modelname in Section~\ref{section:model}, we choose to incorporate $\bm{r}_p$ rather than $g_p$ because a continuous representation can encode more information and reduce sparsity compared to a discrete one.

%% file: sections/s4_model.tex
\section{Goal-directed Web Browsing (GoWeB)}
\label{section:model}



\begin{figure}[!t]
    \centering
    \includegraphics[width=\linewidth]{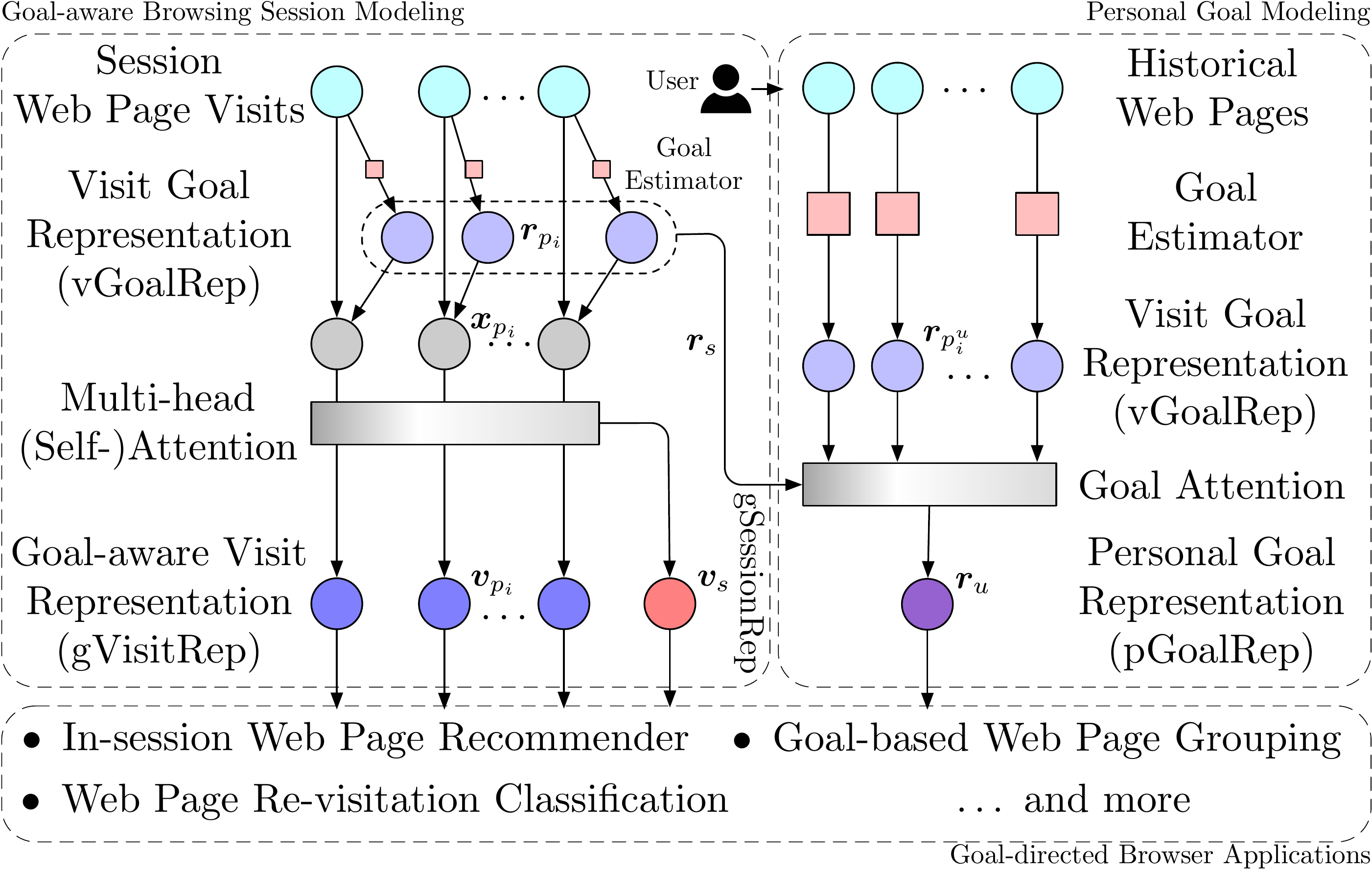}
    
    
    \caption{Overview of \modelname. The goal estimator produces goal representations $r_p$, which are then used to derive goal-aware visit $v_p$ and session $v_s$ representations, as well as personal goal representation $r_u$.
    These representations facilitate goal-directed experiences in browser applications. The details of the goal estimator are described in Section~\ref{section:goalspace}.}
    \label{fig:overall}
\end{figure}

Based on the goal estimator and visit goal representations introduced in Section~\ref{section:goalspace}, we present a unified framework, \modelfullname (\modelname), to assist common activities that people perform when browsing the web. 
We first architect a generic neural framework that can encapsulate information incurred in browsing using goal-aware representations. Second, we employ this framework to assist people to advance their goals through in-session recommendation, to pick up their goals through re-visitation prediction, and to focus on certain goals through goal-based clustering. Figure~\ref{fig:overall} further illustrates an overview of \modelname.




\subsection{Goal-aware Browsing Session Modeling}
\label{section:sessionmodeling}


A web browsing session involves three key actors: individual web pages that a person visits, the sequence of visits in the same session, and the person who performs these visits. To support goal-directed browsing, \modelname needs to derive representations that effectively incorporate goal awareness for each of these. 
More concretely, suppose a web browsing session $s$ consists of $n$ web page visits $s = \left\lbrace p_1, \dots, p_n \right\rbrace$ performed by a person $u$.
In the below, we describe the steps to form individual \textit{goal-aware visit representation} (gVisitRep) as $\bm{v}_{p_i}$, \textit{goal-aware session representation} (gSessionRep) as $\bm{v}_s$, and \textit{personal goal representation} (pGoalRep) as $\bm{r}_u$.

\mysection{Goal-aware Visit Representation (gVisitRep)}
To represent a web page visit, we consider two types of essential signals, namely the lexical content expressed inside the page as well as the probable goals accounting for the visit. 
Specifically, 
we derive content-based embeddings by initializing a separate set of weights using the same encoder architecture described in Section~\ref{section:goalestimator}. 
For each visit $p_i$, the content embedding $\bm{w}_{p_i} \in \mathbb{R}^{d_V}$ is determined by its underlying content $c_{p_i}$ and the host of the page $h_{p_i}$.
Formally, we compute content embedding by 
$ \bm{w}_{p_i} = \mathcal{F}_V([\text{emb}^V_{\text{host}}(h_{p_i}), \text{BERT}^V(c_{p_i})]),$
where $\text{emb}^V_{\text{host}}(h_p)$ projects the host to a $d_h$-dimensional embedding space; $\text{BERT}^V(\cdot)$ is a contextualized language model; $\mathcal{F}_V(\cdot)$ as a fully-connected hidden layer derives the final $d_V$-dimensional content embedding. 
While the content-based embeddings are important, we take a step further to incorporate probable goals for the act of visiting, by concatenating the vGoalRep~$\bm{r}_{p_i}$ and the content embedding $\bm{w}_{p_i}$ as $\bm{x}_{p_i} = [\bm{r}_{p_i}, \bm{w}_{p_i}].$ 
Thus, visits in the same session $s$ can be denoted as $X \in \mathbb{R}^{n\times d_\text{model}}$, where $n$ is the number of pages and $d_\text{model} = d_G + d_V$.


To make page visits sensitive to the session context in which they appear, we propose to apply the multi-head attention mechanism~\cite{vaswani2017attention} to derive contextualized visit representations.
Formally, for each web page visit, the gVisitRep $\bm{v}_{p_i}$ can be computed as:
$$[\bm{v}_{p_1};\dots; \bm{v}_{p_n}] = \text{Concat}(\text{head}_1,\dots,\text{head}_k)W^O,$$
where $\text{head}_i = \text{Attention}(XW_i^Q,XW_i^K,XW_i^V)$; $k$ is the number of attention heads;  $W_i^Q, W_i^K  \in \mathbb{R}^{d_{\text{model}}\times d_k}, W_i^V\in \mathbb{R}^{d_{\text{model}}\times d_v}$, $W^O \in \mathbb{R}^{hd_v \times d_{\text{model}}}$; $d_k$ and $d_v$ are the dimension numbers of attention keys and values; $\text{Attention}(Q,K,V) = \text{softmax}(\frac{QK^T}{\sqrt{d_k}})V$.

\mysection{Goal-aware Session Representation (gSessionRep)}
To represent the overall browsing session, we learn a context vector $\bm{c}_i \in \mathbb{R}^{1\times d_k}$ for each head to dynamically estimate the importance of each visit for a specific application.
The gSessionRep $\bm{v_s}$ of the browsing session can then be derived as
$\bm{v_s} = \text{Concat}(\text{head}^s_1,\dots,\text{head}^s_k)W^{Os},$
where $\text{head}^s_i = \text{Attention}(\bm{c}_i,XW_i^{Ks},XW_i^{Vs})$; $W_i^{Ks}  \in \mathbb{R}^{d_{\text{model}}\times d_k}$, $W_i^{Vs}\in \mathbb{R}^{d_{\text{model}}\times d_v}$, $W^{Os} \in \mathbb{R}^{hd_v \times d_{\text{model}}}.$ 

\mysection{Personal Goal Representation (pGoalRep)}
As an individual's past goals could be predictive of their future goals or behavior, conventionally it is a common practice to rely on an individual's past browsing activities to derive their personal representation~\cite{jiang2020end}.
We argue that it is equally important to account for the activities in a currently active session, according to \citet{shah2002forgetting} that the activation of a given focal goal tends to result in an inhibition of alternative goals. 
To implement this idea, \modelname constructs pGoalRep $\bm{r}_u$ for an individual $u$ by aggregating the current session activities as a query to attend to past activities of the same individual.  
Suppose $\lbrace\bm{r}_{p_i}\rbrace$ denotes the vGoalReps for all page visits in a current session; $Z_u = \lbrace p^u_i \rbrace$ is the set of web pages visited in the past by a user $u$. We adopt the Luong's attention mechanism~\cite{luong2015effective} to aggregate current session activities as:
$$\bm{r}_s = \sum_i \alpha_i \cdot \bm{r}_{p_i}, \alpha_i = \frac{e^{\bm{c}_s\cdot\mathcal{F}_s(\bm{r}_{p_i})}}{\sum_j e^{\bm{c}_s\cdot\mathcal{F}_s(\bm{r}_{p_j})}},$$
where $\alpha_i$ is the weight for $\bm{r}_{p_i}$; $\bm{c}_s$ is the context vector to estimates the importance of each visit; $\mathcal{F}_s(\cdot)$ is a fully-connected hidden layer. 
The pGoalRep $\bm{r}_u$ is computed by using $\bm{r}_s$ as a query to discover and aggregate past goals related to current session:
$$\bm{r}_u = \sum_{p^u_i\in Z_u} \beta_i \cdot \bm{r}_{p^u_i}, \beta_i = \frac{e^{\bm{r}_s \cdot \bm{r}_{p^u_i}}}{\sum_{p^u_j\in Z_u} e^{\bm{r}_s\cdot\bm{r}_{p^u_j}}},$$
where $\beta_i$ is the weight for each historical visit. 

To summarize, we have described how to create goal-aware representations for a page visit ($\bm{v}_{p_i}$), a browsing session ($\bm{v}_s$), and an individual acting in the session ($\bm{r}_u$). These representations can, when designed properly, be used for directing people towards their goals while they browse the web. In particular, we consider three opportunities for goal-aware assistance described as follows.  

\subsection{Task 1: In-session Web Page Recommender}
Making progress towards goals when navigating the web can be challenging, which could be due to unfamiliarity with subject matter or other external factors. To address this, \modelname aims to recommend most goal-related, unseen resources that can help people advance their goals.
Concretely, \modelname formulates an in-session web page recommender. Given $n$ preceding web pages $s=\lbrace p_1, \cdots p_n \rbrace$ visited in a session, \modelname ranks web pages from a candidate set $C$ according to their likelihood of being visited later in the same session.
We follow prior work~\cite{covington2016deep} and cast recommendation as a classification task. 
Based on the information available in $s$, we combine the gSessionRep $\bm{v}_s$ and pGoalRep $\bm{r}_u$ as a feature vector.
The ranking scores $\bm{y}_\text{rec}\in \mathbb{R}^{|C|}$ of each web page in $C$ can then be calculated as classification logits: 
$\bm{y}_\text{rec} = \mathcal{F}_{\text{rec}}(\mathcal{F}_{\text{hidden}}([\bm{v}_s, \bm{r}_u])),$
where $\mathcal{F}_{\text{hidden}}(\cdot)$ is a fully-connected hidden layer; $\mathcal{F}_{\text{rec}}(\cdot)$ projects the hidden state to the ranking scores of candidate web pages.
Finally, the ranked list of recommendations can be generated by retrieving candidates with top ranking scores.

\subsection{Task 2: Re-visitation Prediction}

Web page re-visitation is prevalent as people may visit the same resources for unfinished or recurring goals.
The ability to forecast potential future re-visitations can be useful for supporting people resuming their goals when predicted correctly. 
For instance, the predictions can be 
stored in the browser backend and 
resurface to users when they start a new session as a reminder.  
To support this, \modelname aims to predict whether any of web pages in a current session will be revisited in a future session.
For each visit $p$ in the session $s$, we first construct its feature vector by concatenating the gVisitRep $\bm{v}_p$ and the pGoalRep~$\bm{r}_u$.
The probability of the web page being re-visited can then be estimated as:
$$P(\text{revisit} = \text{True} \mid p, s) = \sigma(\mathcal{F}_{\text{rev}}(\mathcal{F}_{\text{hidden}}([\bm{v}_p,   \bm{r}_u]))),$$
where $\mathcal{F}_{\text{hidden}}(\cdot)$ is a fully-connected hidden layer; $\mathcal{F}_{\text{rev}}(\cdot)$ projects the hidden state to a logit so that the sigmoid function $\sigma(\cdot)$ can derive a probabilistic score.
We note that for the first and second tasks, \modelname is trained end-to-end to directly optimize the target objective.

\subsection{Task 3: Goal-based Web Page Grouping}
\label{section:model:segmentation}
Another common challenge with web browsing is being able to rationalize with goals and focus on subsets of web pages by goals. Considering that people often visit multiple, potentially diverse web pages in a session, this may create high cognitive load when people switch in-between goals. 
For this, we employ \modelname 
to group in-session web pages by goals, such that the resulting groups of pages can be used to categorize ongoing flows and help people concentrate on certain goals of choice. 
Specifically, we cast this as a clustering task where the vGoalRep $\bm{r}_p$ of each visit $p$ is used as features, based on which subsequent feature-based clustering algorithms can be applied.

%% file: sections/s5_experiments.tex
\section{Experiments}
\label{section:exp}

This section examines the effect of incorporating high-ordered goals into the context of web browsing. We compare  each of the browser-centric tasks with competitive baselines and demonstrate how \modelname can enhance multiple browsing experiences.  

\subsection{Experimental Setup}

\mysection{Experimental Datasets} 
Our main experimental dataset, denoted as $\bm{D}_{web}$, consists of web browsing sessions that are used for evaluating the three goal-based applications supported by \modelname.
$\bm{D}_{web}$ was constructed by randomly sampling the anonymized logs of Microsoft Edge web browser\footnote{To the best of our knowledge, unfortunately, there is no publicly available user web browsing dataset that is suitable for evaluating our hypothesis and framework, but we will release our implementation to facilitate the community development.}.
A session is composed of a sequence of web page visits of a user, where a boundary is found between two consecutive visits that are at least 30 minutes apart. 
The logs contain records of web pages visits assembled by sessions, where each page comes with a host, a title and a timestamp when the page was visited; the titles are used as the web page content.
The data was sampled from June 2020, where the training and test sets were respectively gathered from the periods of June 1st to 23rd and June 24th to 30th.
To avoid tail behavior, a web page is discarded if it appears less than 10 times in the sampling period, and short sessions with less than 10 page visits are disregarded as the associated goals are likely to be simpler. 
In total, our dataset contains web pages originating from 79,695 unique hosts (websites).

\input{tables/dataset}

We prepare two disjoint test sets to study the effect of incorporating the notion of personal goals. The warm-start sample includes users whose behavior can be found in the training period, while the cold-start sample draws from users new to the system.
The statistics of $\bm{D}_{web}$ used for the experiments are shown in Table~\ref{tab:dataset}. 
For cold-start users, since we do not have access to their past behavior, we exclude the personal goal  representations from the framework and denote this variant as \textbf{\modelname (NP)}.

\input{tables/recsys}

\mysection{Implementation Details}
We implement \modelname and baseline methods with the PyTorch framework~\cite{paszke2019pytorch}.
The dimensions of goals and hosts (i.e., $d_G$ and $d_h$) are set to 64, while the number of hidden units in fully-connected layers and the dimension of the remaining embeddings are set to 128. 
The number of attention heads $k$ is set to 8.
We use a learning rate of 0.3 for RSGD~\cite{bonnabel2013stochastic} in the Riemannian optimizer, while a learning rate $10^{-5}$ and $(\beta_1, \beta_2) = (0.9, 0.999)$ are used in the Adam optimizer~\cite{kingma2014adam} for optimizing the supervised goal-based applications and the goal estimator. 
For contextualized language modeling (i.e., the $\text{BERT}$ functions in Section \ref{section:model} and \ref{section:goalspace}), we use the pre-trained RoBERTa~\cite{liu2019roberta} models provided by HuggingFace~\cite{wolf2019huggingface}.
We will open-source our implementation.

\subsection{Web Page Recommendation Performance}
Recall that the in-session recommender predicts goal-related web pages that are likely to be visited later in the same session.
Specifically, for a session, we take the first half of page visits as observed input information
and use the second half as ground-truth to evaluate our model.
The ranking candidates of web pages are generated by removing the top 10 popular web pages (e.g., \texttt{msn.com} and \texttt{google.com}) and selecting the top 50,000 frequent web pages\footnote{The number was selected in accordance with the prior study ~\cite{Adar2008revisit} that analyzed large-scale web data.} from the remaining set.
For both test sets, we assert that the ground-truth (i.e., the second half) of every session contains at least one web page from the candidate set.
We treat the task as a retrieval problem and evaluate it by conventional metrics, including MRR@10, HR@\{1, 5, 10\}, and NDCG@\{5, 10\}.
We compare our method with state-of-the-art top-$K$ session-based recommenders, including popularity~(Pop), BPRMF~\cite{Rendle2009BPR}, NCF~\cite{He2017NeuralCF}, Caser~\cite{Tang2018PersonalizedTS}, {GRU4Rec}~\cite{Hidasi2016SessionbasedRW}, {SASRec}~\cite{Kang2018SelfAttentiveSR}, and {BERT4Rec}~\cite{Sun2019BERT4Rec}.
We further consider a modern content-based modeling method that incorporates web page semantics using BERT and contextualization. We implement this by reusing the \modelname neural architecture but with all the goal-awareness removed (i.e. excluding $r_p$ and any representations derived based on $r_p$), which is denoted as SemRec.   

Table~\ref{tab:recsys} demonstrates the recommendation performance in two test sets.
Among the baseline methods, SemRec performs better than conventional session-based recommenders that disregard web page content.
It advocates for the importance of modeling the semantics of web pages, as the open-domain nature of the web can make recommendation more challenging than domain-specific applications (e.g. movies).
Without personalization, \modelname (NP) performs the best and outperforms SemRec by 18.52\% and 16.09\% in MRR@10 for two test sets. 
This empirical finding sheds light on the opportunity of modeling human motives and higher-level goals carried by the goal estimator without accessing historical data, since only the in-session page visits are considered. 
For the warm-start test set where access to past page visits is available, \modelname can further provide more satisfactory recommendations.
Overall, the results suggest that incorporating people's motives can lead to better recommendation in web browsing sessions. 

\subsection{Re-visitation Prediction Performance}

For re-visitation prediction, given a browsing session, the task is formulated as a binary classification problem to predict if each web page visit in the session will be re-visited by the same user in any future session.
Similarly, we consider two families of baseline methods.
The conventional sequence modeling methods learn an embedding vector for each unique web page without using web page content, including CNN, RNN, SAS~\cite{Kang2018SelfAttentiveSR}, BERT~\cite{devlin2019bert,liu2019roberta}.
To account for page content, we likewise build a strong semantic classifier, SemCLS, based on the sequence modeling component embedded in \modelname without using any goal-related representations.
We utilize common binary classification metrics, including F1-score, precision, recall, and accuracy, for evaluation.

\input{tables/revisit}

Table~\ref{tab:revisit} shows the classification performance of methods in two test sets.
Consistent with the results in Table~\ref{tab:recsys}, SemCLS that incorporates web page content is the best-performing baseline method.
Compared to SemCLS, \modelname(NP) respectively provides 2.32\% and 3.30\% gains in F1 scores for the two test sets.
Similar to in-session recommender, \modelname introduces further gains when using the focal goals to attend to past goals.
These results indicate that the \modelname framework can provide goal-aware assistance in supervised ranking and classification tasks.

\subsection{Goal-based Grouping Performance}

The objective of the web page grouping task is to create coherent clusters of web pages inside a session in an unsupervised fashion.
We treat the goal representations $r_p$ derived in \modelname as  
features and apply $K$-means++ algorithm~\cite{arthur2006k} to cluster page visits into different groups.
We consider two popular text modeling methods, doc2vec~\cite{le2014distributed} and BERT~\cite{devlin2019bert}, as the baselines where the corresponding embedding vectors are used for clustering.

To create ground-truth for evaluation, we leverage the outcome of a proprietary hierarchical topical classifier as references that different methods can compare to.
The proprietary classifier predicts the topics of web pages according to a 3-layer taxonomy consisting of a root node, 22 category nodes, and 288 leaf nodes.
This setup resembles the traditional web page classification task \cite{Baykan2011ACS, Qi2009WebPC} where a topical organization of the web was found to be useful. 
Once we obtain the outcome topics, the number of ground-truth clusters in each session is also determined accordingly. 
Note that, following previous studies~\cite{yang2019deep,law2017deep}, our focus is to evaluate the clustering quality using different embedding features. We acknowledge past work on determining the optimal number of clusters~\cite{kodinariya2013review} as a different research problem.

\input{tables/clustering}

\begin{figure}[!t]
    \centering
    \includegraphics[width=\linewidth]{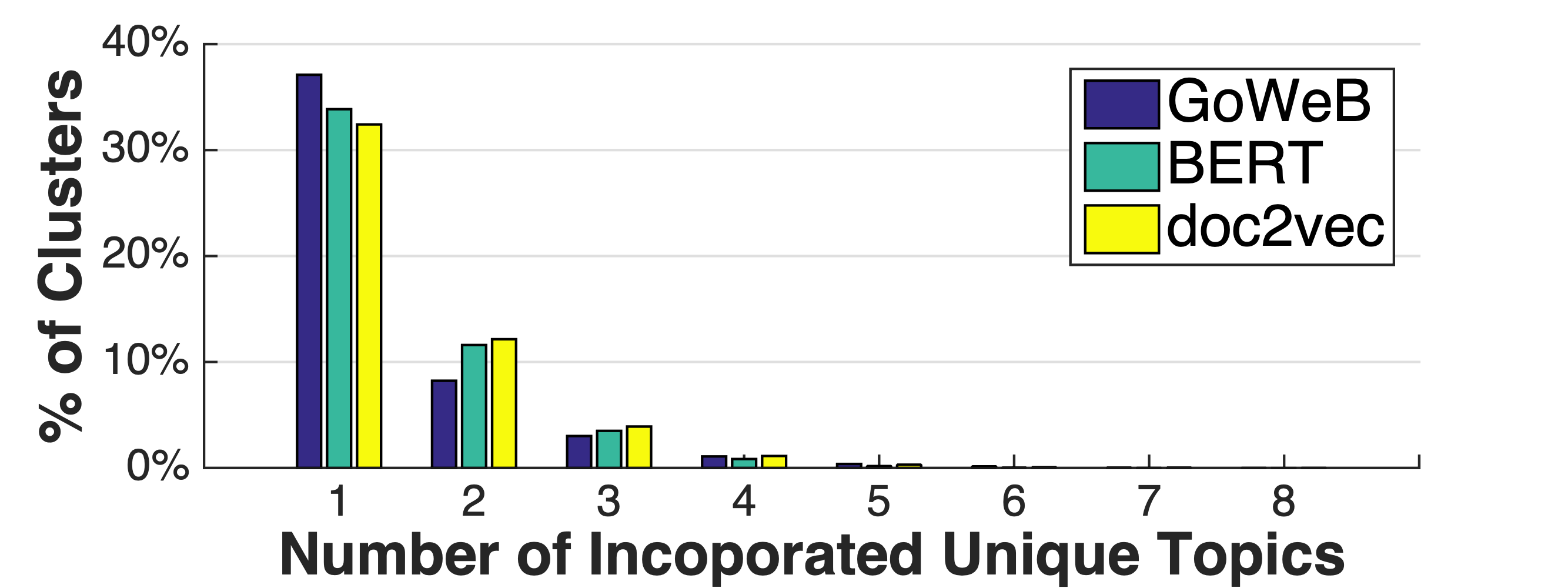}
    \caption{The percentage of clusters over different numbers of incorporated unique topics.}
    \label{fig:goal_density}
\end{figure}

Table~\ref{tab:clustering} shows the resulting grouping performance on NMI and AMI metrics~\cite{manning2008introduction}; here we combine the two test sets together since this is an unsupervised task.
BERT, as is powered by a decent pretrained contextualized language model, performs better than doc2vec. 
Meanwhile, the goal representations generated by goal estimatoroutperforms both baselines, suggesting its efficacy for tackling unsupervised  tasks. Figure~\ref{fig:goal_density} further shows the percentage of clusters over the number of unique ground-truth topics included in a cluster. An ideal cluster should contain exactly one topic.

The results show that \modelname tends to derive purer clusters with fewer unique classes compared to baselines.
It may suggest that the goal-aware representations are also more topically coherent.

%% file: tables/dataset.tex
\begin{table}[!t]
    \centering
    \renewcommand{\arraystretch}{0.9}
    
    \resizebox{\linewidth}{!}{
    \begin{tabular}{c|ccc} \hline
    & \# of Users & \# Sessions & Avg. Session Length \\ \hline
Training & 57,715 & 836,073 & 21.689 $\pm$ 11.922 \\ 
Test (Warm) & 6,701 & 38,737 & 22.257	$\pm$ 12.124 \\ 
Test (Cold) & 6,062 & 75,339 & 22.076 $\pm$ 12.127 \\ \hline
    \end{tabular}}
    \caption{Statistics of web session experimental datasets.}
    \label{tab:dataset}
    \vspace{-12pt}
\end{table}

%% file: tables/recsys.tex
\begin{table*}[!t]
    \centering
    \resizebox{\linewidth}{!}{
    \begin{tabular}{c|c|cccccccc|c|c} \hline
 &  & Pop & BPRMF & NCF & Caser & GRU4Rec & SASRec & BERT4Rec & SemRec & GoWeB (NP) & GoWeB \\ \hline
 & MRR@10 & 0.0386 & 0.0484 & 0.0537 & 0.0506 & 0.0533 & 0.0693 & 0.0727 & 0.0817 & 0.0968 (+18.52\%)& \bf 0.1080 	(+32.22\%)\\
 & HR@1 & 0.0169 & 0.0243 & 0.0259 & 0.0216 & 0.0269 & 0.0365 & 0.0382 & 0.0472 & 0.0581 (+23.24\%)
&  \bf 0.0674	(+42.96\%) \\
Test & HR@5 & 0.0618 & 0.0751 & 0.0864 & 0.0870 & 0.0837 & 0.1093 & 0.1150 & 0.1252 & 0.1445 (+15.38\%) &  \bf 0.1585 	(+26.53\%)\\
(Warm) & HR@10 & 0.1123 & 0.1243 & 0.1365 & 0.1329 & 0.1445 & 0.1681 & 0.1748 & 0.1801 & 0.2090 (+16.05\%)	&  \bf 0.2276 (+26.38\%)\\
 & NDCG@5 & 0.0202 & 0.0276 & 0.0295 & 0.0278 & 0.0303 & 0.0398 & 0.0418 & 0.0476 & 0.0563 (+18.13\%) &  \bf 0.0626 	(+31.34\%)\\
 & NDCG@10 & 0.0292 & 0.0354 & 0.0373 & 0.0363 & 0.0406 & 0.0503 & 0.0521 & 0.0571 & 0.0676 (+18.33\%)&  \bf 0.0751 	(+31.43\%)\\ \hline
 & MRR@10 & 0.0454 & 0.0513 & 0.0528 & 0.0508 & 0.0559 & 0.0675 & 0.0683 & 0.0856 & \bf 0.0993 (+16.09\%) & N/A \\
 & HR@1 & 0.0215 & 0.0234 & 0.0268 & 0.0251 & 0.0267 & 0.0353 & 0.0337 & 0.0499 	& \bf  0.0597 (+19.66\%) & N/A \\
Test & HR@5 & 0.0752 & 0.0882 & 0.0884 & 0.0829 & 0.0937 & 0.1034 & 0.1094 & 0.1350 &  \bf 0.1496 (+10.83\%)	& N/A \\
(Cold) & HR@10 & 0.1206 & 0.1355 & 0.1269 & 0.1340 & 0.1493 & 0.1696 & 0.1761 & 0.1874 &  \bf 0.2150 (+14.68\%)	& N/A \\
 & NDCG@5 & 0.0252 & 0.0318 & 0.0316 & 0.0280 & 0.0335 & 0.0392 & 0.0405 & 0.0518 &  \bf 0.0598 (+15.30\%)	 & N/A \\
 & NDCG@10 & 0.0341 & 0.0398 & 0.0383 & 0.0375 & 0.0435 & 0.0505 & 0.0520 & 0.0611 &  \bf 0.0712 (+16.50\%)	& N/A \\ \hline
    \end{tabular}}
    \caption{Performance of methods in session-based web page recommendation. Note that cold-start users do not have personalized historical goal embeddings from training data, so \modelname  \new{with personalization} is not available for the cold-start testing dataset. \new{\modelname (NP) denotes  the non-personalized version of \modelname.}}
    \label{tab:recsys}
    \vspace{-18pt}

\end{table*}

%% file: tables/revisit.tex
\begin{table}[!t]
\centering
\resizebox{\linewidth}{!}{
\begin{tabular}{c|c|cccc} \hline
 &  & F1 & Precision & Recall & Accuracy \\ \hline
 & CNN & 0.5249 & 0.4818 & 0.5765 & 0.6066 \\
 & RNN & 0.5387 & 0.6269 & 0.4722 & 0.6951 \\
\multirow{2}{*}{Test} & SAS & 0.5302 & 0.6376 & 0.4537 & 0.6969 \\
\multirow{2}{*}{(Warm)} & BERT & 0.5948 & 0.5479 & 0.6505 & 0.6659 \\
 & SemCLS & 0.6718 & 0.6155 & 0.7394 & 0.7277 \\ \cline{2-6}
 & GoWeB (NP) & 0.6874 & 0.6247 & 0.7642 & 0.7380 \\
 & GoWeB & \bf 0.7010 & \bf 0.6234 & \bf 0.8007 & \bf 0.7425 \\ \hline
 & CNN & 0.5454 & 0.5271 & 0.5650 & 0.6075 \\
 & RNN & 0.5369 & 0.6826 & 0.4425 & 0.6819 \\
Test & SAS & 0.5290 & 0.6956 & 0.4268 & 0.6833 \\
(Cold) & BERT & 0.6127 & 0.6093 & 0.6160 & 0.6754 \\
 & SemCLS & 0.6824 & 0.6829 & 0.6820 & 0.7355 \\ \cline{2-6}
 & GoWeB (NP) & \bf 0.7050 & \bf 0.6984 & \bf 0.7117 & \bf 0.7518 \\\hline
\end{tabular}}
\caption{Performance of methods in re-visitation classification. Similar to Table~\ref{tab:recsys}, \modelname \new{with personalization} is N/A for cold-start users. \new{\modelname (NP) denotes the non-personalized version of \modelname.}}
    \label{tab:revisit}
\end{table}

%% file: tables/clustering.tex
\begin{table}[!t]
    \centering
    \resizebox{\linewidth}{!}{
    \begin{tabular}{c|cccc} \hline
  & doc2vec & BERT & \modelname (Goal Estimator) \\ \hline
NMI & 0.7290 & 0.7612 & \bf 0.7683 (+0.93\%) \\
AMI & 0.4320 & 0.4929 & \bf 0.5155 (+4.59\%) \\ \hline
    \end{tabular}}
    \caption{Performance of methods in goal-based clustering.}
    \label{tab:clustering}
\end{table}

%% file: sections/s6_analysis.tex
\section{Analysis and Discussions}
\label{section:analysis}



In section~\ref{section:exp}, we demonstrate empirically the effectiveness of \modelname when applying to browser applications.
Now we turn our attention to deepen our understanding for the effectiveness of the goal estimator, and how the variety of goals may affect the differences observed in people's behavioral patterns.

\begin{figure}[!t]
\includegraphics[width=\linewidth]{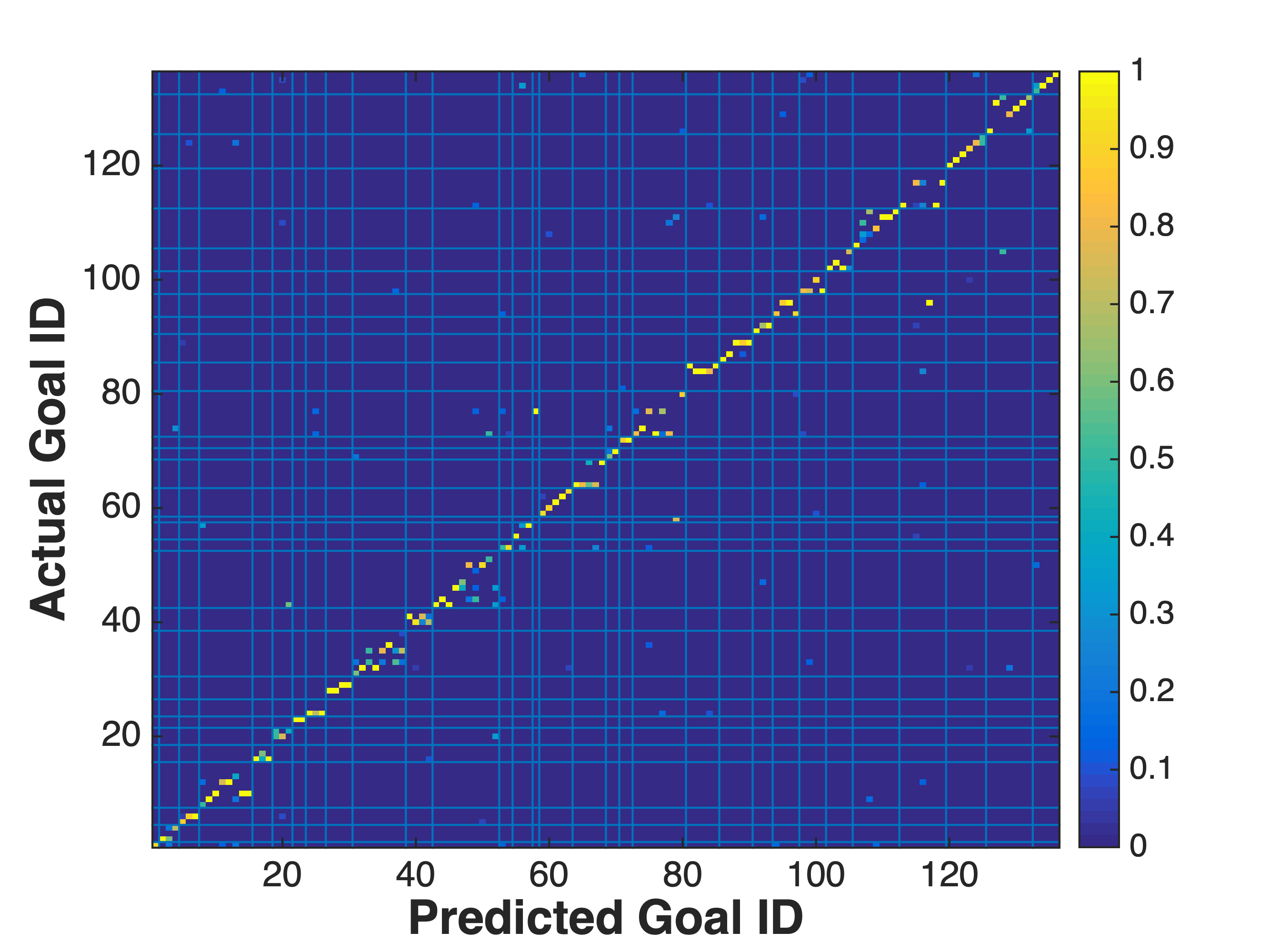}
\captionof{figure}{The probabilistic confusion matrix for our goal estimator in goal classification. The grid partitions goals into categories. The ``non-goal'' class has goal ID 1.}%
\label{fig:goal_heat_map}
\end{figure}

\mysection{Effectiveness of Goal Estimator}
To uncover the details, we scrutinize the goal estimator as it powers the prediction for 
goal representations when people make visits to web pages. 
Recall that the goal estimator is learned with a multi-class objective, where a 136-way classification is carried out to predict among the 135 leaf goals plus a non-goal class. 
We use 90\% of weakly-labeled instances from $\bm{D}_{weak}$ for training, and evaluate on the remaining 10\%.
As a result, we achieve 55.27\% and 83.01\% F1 scores for predicting individual goals and goal categories.
Figure~\ref{fig:goal_heat_map} depicts the confusion matrix for our goal estimator, where the grid partitions leaf goals into goal categories.
We can see that most of the predictions are aligned with the diagonal as correct predictions; among other cases, most of the misclassifications are in the same grid block, indicating that errors are bounded within the same goal category.

\begin{figure*}[!t]
    \begin{subfigure}[t]{.49\linewidth}
        \centering
        \includegraphics[width=\linewidth]{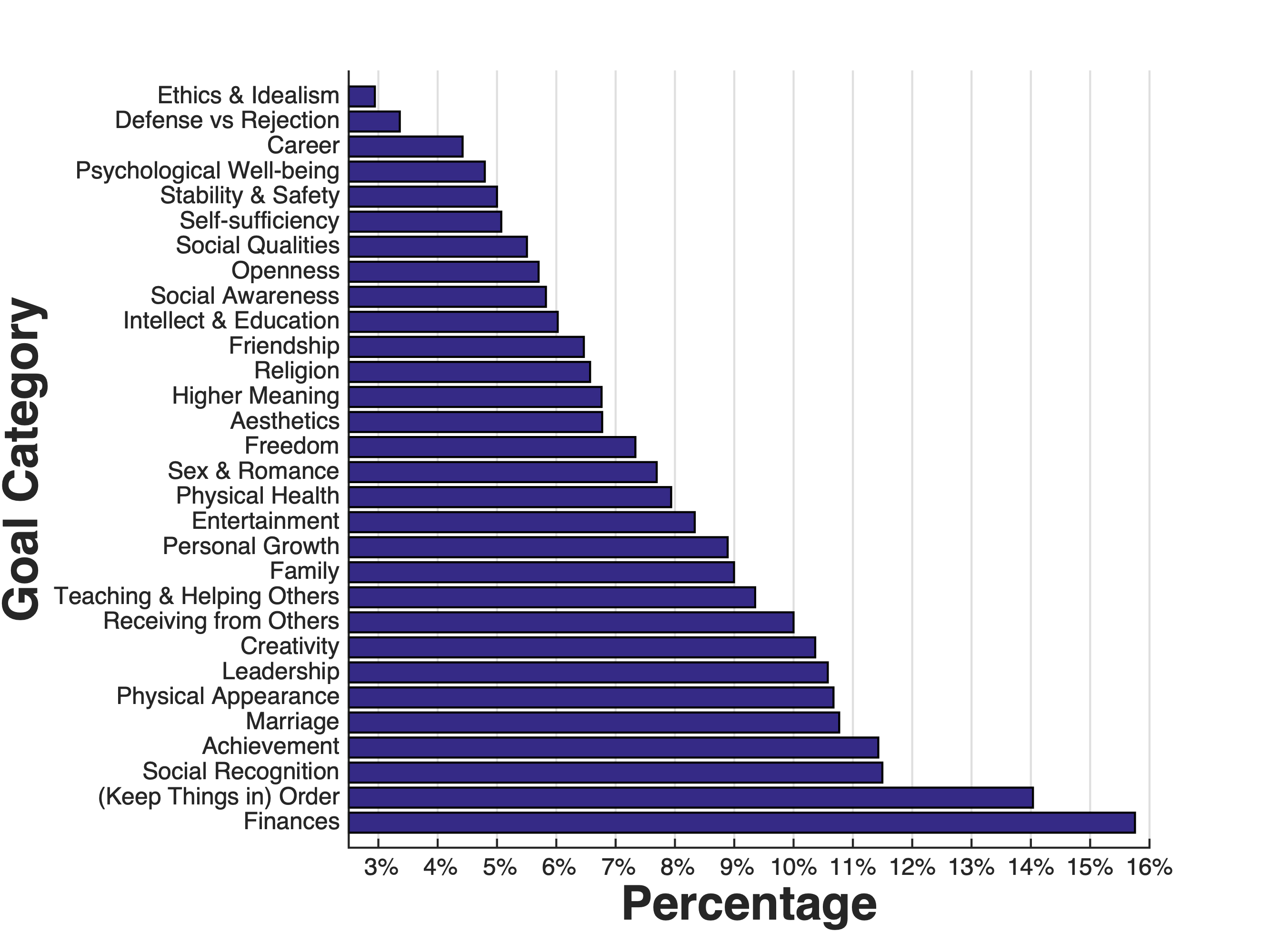}
        \caption{}
        \label{fig:cogoal}
    \end{subfigure}
    \begin{subfigure}[t]{.49\linewidth}
        \centering
        \includegraphics[width=\linewidth]{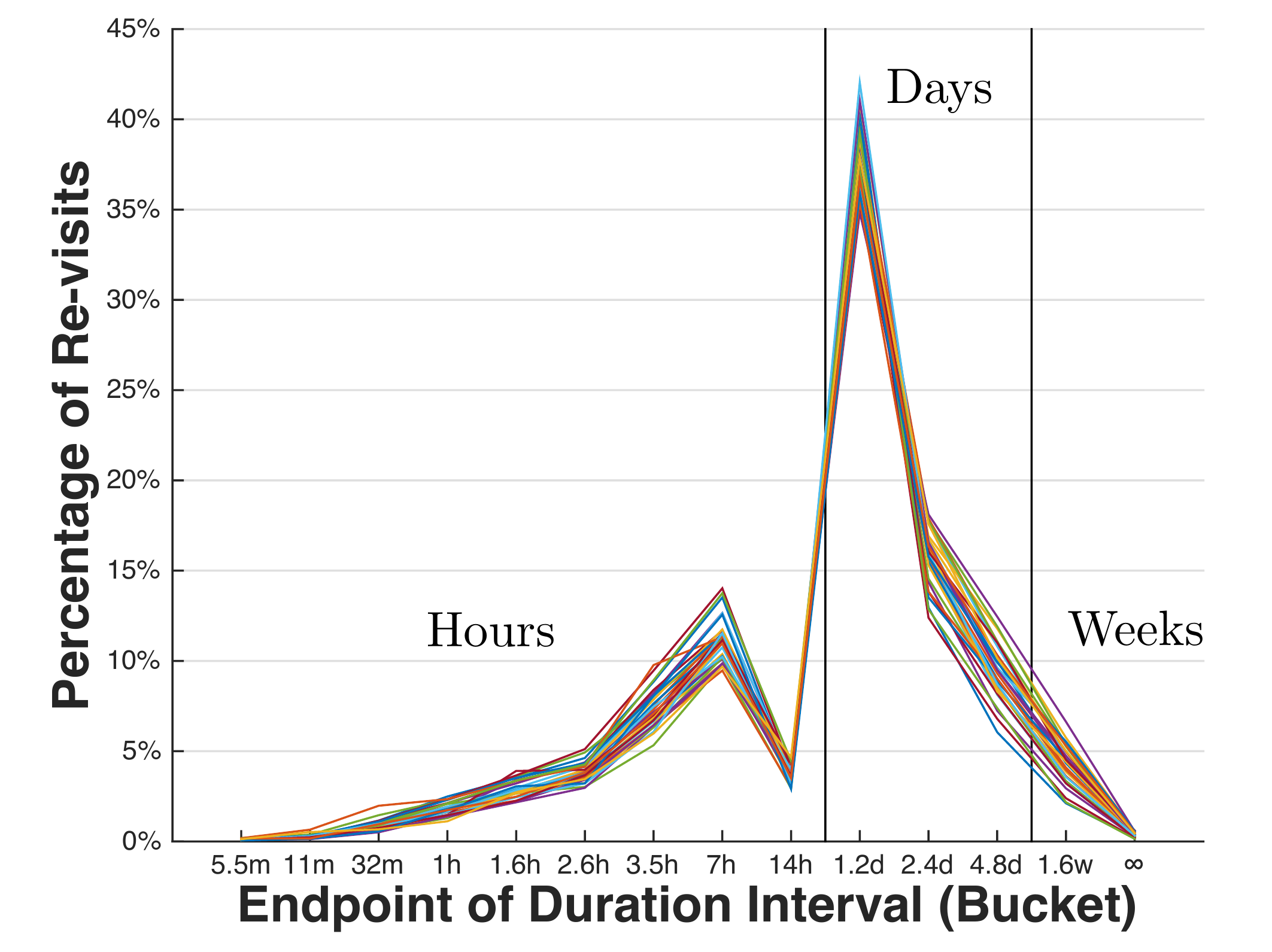}
        \caption{}
        \label{fig:duration}
    \end{subfigure}
    
    \caption{(a) The percentage of being in a web browsing session with a single goal for each goal category. (b) Percentage of re-visits within different duration interval buckets~\cite{Adar2008revisit} and three duration scales for goal categories. Note that each line represents the statistics of goals in a certain category.}
\end{figure*}


We further conduct a lightweight qualitative exercise to analyze the relations between hosts and predicted goals. 
We first identify top hosts that cover a higher proportion of web pages being predicted closest to each goal category.
Two annotators were then asked to judge whether these hosts are topically relevant to the corresponding categories. 
Table~\ref{tab:tophost} shows the assessment results with 100\% agreement from the two annotators.
The assessments suggest that the goal estimator can make reasonable predictions in the majority of cases.  

\input{tables/tophost}



\mysection{Characterization of In-session Goals}
To start, we note that people tend to involve in multiple goals in a single session; for example, the average number of goals can be more than 5 for sessions containing 10 page visits. 
We also find that the increase in the number of goals pursued is sub-linear to the session length; for example, sessions of 60 page visits are associated with fewer than 14 goals on average.
Given that people may seek multiple goals in a session, to which degree each goal category could be considered as the session's focal goal?
Figure~\ref{fig:cogoal} illustrates the percentage of being the only goal in a single session. 
On one hand, for sessions that contain a \textit{Finances} goal, about 16\% of the times people concentrate solely on this goal category. We conjecture that it might be related to the time-dependent sensitivity required in the process of financial decision making; further studies need to be conducted in the future.
On the other hand, we find that more nebulous goals such as \textit{Ethics \& Idealism} are seldom being pursued singularly (i.e., only 3\% of the times).
The results suggest people behave differently according to the types of goals, and fathoming the underlying objectives is key to providing delightful web browsing experiences. 
For example, enabling a focus mode on detecting goals that require high concentration, or proactively recommending new resource for more exploratory goals could be new designs to consider.


\mysection{Characterization of Revisitation Patterns by Goals}
People's goals that motivate visits to certain web pages may also have an impact on how these pages are revisited. 
We analyze temporal cross-session revisitation patterns. 
A revisit is identified by two consecutive visits of the same web page, where the elapsed time between the former visit and the latter is considered as the duration.
Following~\citet{Adar2008revisit}, we group all collected revisitation patterns into 14 unequal-size buckets according to the length of duration.
Figure~\ref{fig:duration} illustrates the distributions of revisits in different buckets for different goal categories.
Interestingly, at a high level, all the 30 goal categories exhibit similar revisiting patterns.
Most of the revisits (37\%) occur roughly after a day or within few days, while between 23\% to 36\% of revisits occur within 7 hours.
We observe far less revisiting after a week has passed. This could be due to the limitation of our data being sampled from a month period. 

\input{tables/topduration}
Beyond the distributional similarity at a macro level, we observe distinct patterns in the top 5 goal categories for three revistation duration scales in Table~\ref{tab:topduration}.
The results suggest that goal categories concerning social interactions (\textit{Friendship}, \textit{Social Recognition,} and \textit{Marriage}), addictability (\textit{Entertainment}), and timeliness (\textit{Appearance}) can lead to relatively quicker revisits within hours.
It may imply that inter-personal relationships tend to urge people to voluntarily stay up-to-date, potentially due to the desire for intimacy or attractive looks, or the fear of missing out.
On the contrary, goal categories (\textit{Career} development, \textit{Self-sufficiency,} and \textit{Social Qualities}) that demand continuous investment associate more frequently with slower revisits. 
More abstract, higher-ordered goals, such as \textit{Order,} \textit{Receiving from Others} and \textit{Stability \& Safety} may result in the slowest revisitation compared to more concrete goals such as \textit{Entertainment}.
This highlights the opportunities for supporting people browsing the web, where an ambient reminder for a potential revisit can reduce the overhead of manual retrieval.



%% file: tables/tophost.tex
\begin{table}[!t]
    \centering
    \resizebox{\linewidth}{!}{
    \begin{tabular}{|c|c|c|c|}\hline
        Goal Category & Top 1 & Top 2 & Top 3  \\ \hline
Appearance &  alphacute.com (\checkmark) &  interesticle.com (\checkmark)&  trend-chaser.com (\checkmark)  \\
Career &  indeed.com (\checkmark)&   careerbuilder.com (\checkmark) &   linkedin.com (\checkmark)\\
Entertainment & en.wikipedia.org &  gocomics.com (\checkmark)& nytimes.com \\
Finances &   cnbc.com (\checkmark)&  zillow.com (\checkmark) &   realtor.com (\checkmark) \\
Order &  microsoft.com  (\checkmark) & indeed.com &  lowes.com  (\checkmark) \\
Physical Health &   healthygeorge.com (\checkmark)  &   myfitnesspal.com  (\checkmark) & financialadvisorheroes.com \\
Religion &   biblegateway.com  (\checkmark) &   kingjamesbibleonline.org  (\checkmark) &   bibletrivia.com  (\checkmark) \\ \hline
    \end{tabular}}
    \caption{Top 3 hosts whose web pages are more likely to belong to some goal categories. Hosts with (\checkmark) are labeled as relevant websites to the corresponding categories. The goal ``Order'' means ``(Keep Things in) Order.''}  
    \label{tab:tophost}
\end{table}

%% file: tables/topduration.tex
\begin{table}[!t]
    \centering
    \renewcommand{\arraystretch}{0.9}
    \resizebox{\linewidth}{!}{
    \begin{tabular}{c|c|c|c} \hline
        \multirow{2}{*}{Rank}     & \multicolumn{3}{c}{Duration Scale} \\\cline{2-4} 
         & Hours & Days & Weeks \\ \hline
        1 & Friendship &  Romance & (Keep Things in) Order \\ 
        2 & Appearance & Social Qualities  & Receiving from Others \\
        3 & Entertainment & Career & Career \\ 
        4 & Social Recognition & Self-sufficiency & Stability \& Safety \\
        5 & Marriage & Religion & Social Qualities \\ \hline
    \end{tabular}}
    \caption{Top 5 goal categories with different duration scales in re-visitations.}
    \label{tab:topduration}
    \vspace{-12pt}
\end{table}

%% file: sections/s6_conclusion.tex

\section{Conclusion}
\label{section:conclusion}

In this paper, we highlight the importance of modeling human motives grounded by the long history of psychology literature. A unified neural framework, \modelname, is presented to fulfill this vision.
We build on the top of existing taxonomy and concertize these goals with structure-preserving representation learning in hyperbolic space, based on which a goal estimator is introduced to tighten the loop of how goals could be employed for enhancing browsing experiences. 
We showcase the generality of \modelname and adopt it in three browser-centric applications. 
On real-world data, \modelname consistently outperforms competitive baselines for both warm-start and cold-start users, and demonstrates additional gains when using the focal goals to attend to past goals.
Our follow-up analysis reveals the effectiveness of the goal estimator via quantitative and qualitative exercises, and characterizes the similarities and differences found in behavioral patterns when people pursue different goals. 

Our work brings new perspectives in multiple ways. We present a promising paradigm where we capture the fundamental motives that drive people in their actions and reflect those in digital applications. Broadly speaking, we introduce and transfer the knowledge from psychology findings to modeling browsing sessions on the web, while keeping the framework flexible such that future research could incorporate other types of taxonomies.  
It is our hope that these findings can lift the burden of understanding and characterizing complex human goals for ubiquitous web browsing applications.


